\documentclass{article}

\usepackage{arxiv}

\usepackage[utf8]{inputenc} % allow utf-8 input
\usepackage[T1]{fontenc}    % use 8-bit T1 fonts
\usepackage{hyperref}       % hyperlinks
\usepackage{url}            % simple URL typesetting
\usepackage{booktabs}       % professional-quality tables
\usepackage{amsfonts}       % blackboard math symbols
\usepackage{nicefrac}       % compact symbols for 1/2, etc.
\usepackage{microtype}      % microtypography
\usepackage{lipsum}
\usepackage{color}
\usepackage{algorithm}
\usepackage{algorithmic}
\usepackage{amsmath,amsfonts,amssymb}
\usepackage{subfigure}
\usepackage{lineno}
\usepackage{graphicx}
\usepackage{bm}
\usepackage{subfigure}

\title{Single-snapshot machine learning\\
for super-resolution of turbulence}

\author{
Kai Fukami$^{*}$ and Kunihiko Taira
\\
\\
Department of Mechanical and Aerospace Engineering\\
University of California, Los Angeles, CA 90095, USA\\
$^{*}$Corresponding author: kfukami1@g.ucla.edu
}

\begin{document}
\maketitle

\begin{abstract}

\vspace{-2mm}
Modern machine-learning techniques are generally considered data-hungry.
However, this may not be the case for turbulence as each of its snapshots can hold more information than a single data file in general machine-learning settings. 
This study asks the question of whether nonlinear machine-learning techniques can effectively extract physical insights even from as little as a {\it single} snapshot of turbulent flow.
As an example, we consider machine-learning-based super-resolution analysis that reconstructs a high-resolution field from low-resolution data {for two examples of two-dimensional isotropic turbulence and three-dimensional turbulent channel flow.
First,} we reveal that a carefully designed machine-learning model trained with flow tiles sampled from only a single snapshot can reconstruct vortical structures across a range of Reynolds numbers for two-dimensional decaying turbulence.
Successful flow reconstruction indicates that nonlinear machine-learning techniques can leverage scale-invariance properties to learn turbulent flows.
We also show that training data of turbulent flows can be cleverly collected from a single snapshot by considering characteristics of rotation and shear tensors.
{Second, we perform the single-snapshot super-resolution analysis for turbulent channel flow, showing that it is possible to extract physical insights from a single flow snapshot even with inhomogeneity.}
The present findings suggest that embedding prior knowledge in designing a model and collecting data is important for a range of data-driven analyses for turbulent flows.
More broadly, this work hopes to stop machine-learning practitioners from being wasteful with turbulent flow data.
\end{abstract}

\section{Introduction}
\label{sec:intro}

By gazing at a turbulent flow acquired from numerical simulation or experiment, we can admire the rich physics that involves swirling, stretching, and diffusion.
Turbulence also presents multi-scale characteristics over broad length scales~\cite{davidson2015turbulence}.
In high Reynolds number turbulent flows, the rich phenomena and characteristics are exhibited at any instance in time.
We argue that even a single snapshot of turbulent flow can hold sufficient information to train machine-learning models.
This paper poses a question of whether a commonly used big data set is required for training machine-learning models in studying turbulence.

There have been increased usages of modern machine-learning techniques to analyze, model, estimate, and control turbulent flows~\cite{BNK2020}.
These applications include subgrid-scale modeling~\cite{DIX2019}, reduced-order modeling~\cite{racca2023predicting}, super resolution/flow reconstruction~\cite{FFT2021,guastoni2021convolutional,cuellar2024three}, and flow control~\cite{duriez2017machine,park2020machine}.
These machine-learning models require enormous amount of training data, which is generally significantly larger than those necessitated by traditional analysis techniques.

However, it may be possible to extract important flow features without such large data sets since even a single turbulent flow snapshot contains multi-scale, scale-invariant structures.
To achieve meaningful learning from a single snapshot, we consider training machine-learning models through subsampling and leveraging turbulent statistics.
We further note that it is important that machine-learning models have appropriate architectures and learning formulation that fold in physics~\cite{raissi2019physics,BK2019,lee2019data,FFT2023_survey}.

This study considers data-driven analysis using only a single training snapshot of turbulent flow.
As {examples}, we perform machine-learning-based super-resolution analysis for two-dimensional decaying turbulence {and three-dimensional turbulent channel flow}.
We show that flow reconstruction over a range of Reynolds numbers is possible with nonlinear machine learning by cleverly sampling data from a single snapshot.
The present results show that a large data set is not necessarily needed for machine learning of turbulent flows.

This paper is organized as follows.
The approach is described in section~\ref{sec:method}.
Results from the single-snapshot super-resolution analysis are presented in section~\ref{sec:res}. 
Conclusions are offered in section~\ref{sec:conc}.

% \vspace{-6mm}
\section{Approach}
\label{sec:method}

{
The objective of this study is to show that it is possible to perform data-driven analysis of turbulent flows with a very limited amount of training data -- even from a single snapshot.
For the present analysis, we consider machine-learning-based super-resolution reconstruction of fluid flows~\cite{FFT2019a}.
A machine-learning model $\cal F$ is trained to reconstruct a high-resolution flow field ${\bm q}_{\rm HR}$ from a low-resolution data ${\bm q}_{\rm LR}$:
\begin{align}
{\bm q}_{\rm HR} = {\cal F}({\bm q}_{\rm LR};{\bm w}),    
\end{align}
where $\bm w$ denotes the weights inside the model. 
In this study, the model $\cal F$ is trained with a collection of subdomains sampled from only {\it a single snapshot} of two-dimensional isotropic turbulence {and three-dimensional turbulent channel flow}.
The model is then tested with independent snapshots.
If the model $\cal F$ successfully learns the relationship between low- and high-resolution flow fields from a single training snapshot, we expect that the reconstruction would be possible even for independent testing conditions.
}

\begin{figure}
    \centering
    \includegraphics[width=1\textwidth]{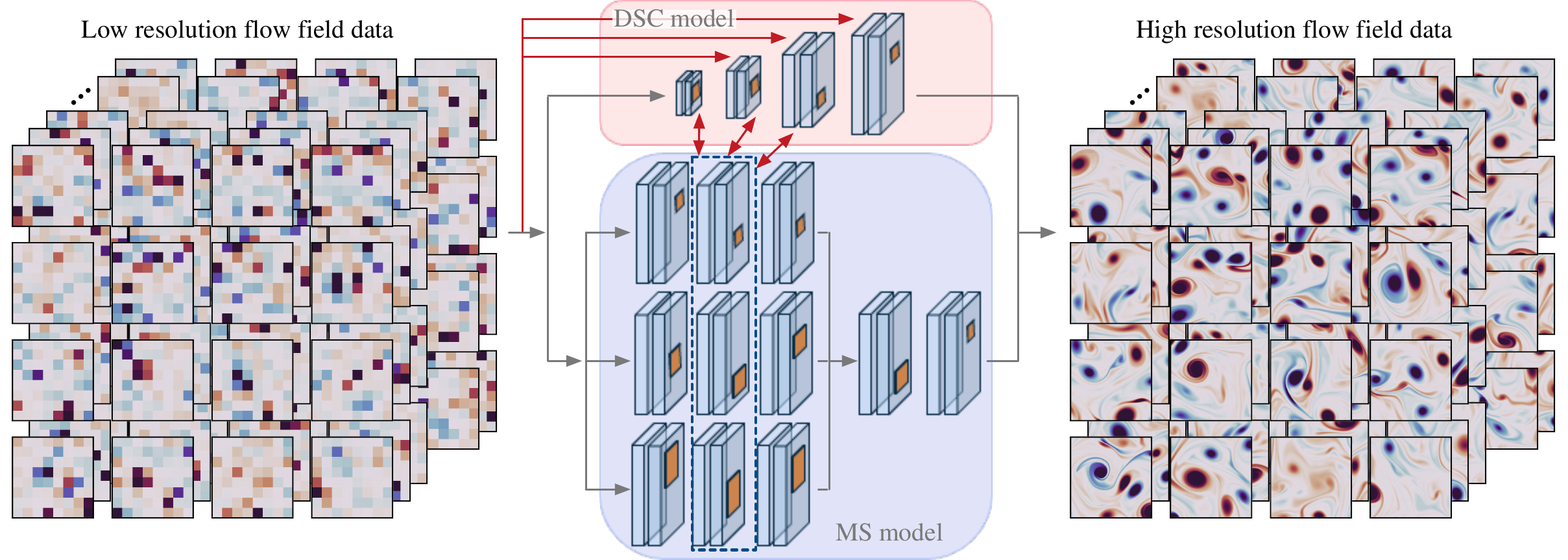}
    % \vspace{-5mm}
    \caption{
    Interconnected DSC/MS model~\cite{FFT2023_survey} for super-resolution reconstruction of turbulent flows.
    }
    % \vspace{-3mm}
    \label{fig2}
\end{figure}

{
For machine-learning-based super resolution of turbulent flows, the model ${\cal F}$ needs to be carefully designed to accommodate a range of length scales while accounting for rotational and translational invariance of vortical structures~\cite{FFT2021}.
This study uses the interconnected hybrid downsampled skip-connection/multi-scale (DSC/MS) model~\cite{FFT2023_survey} based on convolutional neural networks (CNN)~\cite{LBBH1998}, as illustrated in figure~\ref{fig2}.
Between the layers $(l-1)$ and $(l)$, the CNN learns the nonlinear relationship between input and output data by extracting spatial features of given data through filtering operations,
\begin{equation}
    c^{(l)}_{ijn}=\varphi\left(\sum_{m=1}^M\sum_{p=0}^{H-1}\sum_{q=0}^{H-1}h^{(l)}_{pqmn}c^{(l-1)}_{i+p-G,j+q-G,m}+b_n^{(l)}\right),
    \label{eq:CNN}
\end{equation}
where $G=\lfloor H/2\rfloor$, $H$ is the width and height of the filter $h$, $M$ is the number of input channel, $n$ is the number of output channel, $b$ is the bias, and $\varphi$ is the activation function.
By using a nonlinear function for $\varphi$, the convolutional networks can account for nonlinearlities in learning features from training data.
}

The DSC model (boxed in red) includes up/downsampling operations and skip connections, capturing rotational and translational invariance~\cite{FGT2024}.
The MS model (boxed in blue) consists of three different sizes of filter operations, enabling the model to learn a range of length scales in turbulent flows.
Furthermore, these two networks are internally connected via skip connections~\cite{he2016deep} to enhance the correlation of the intermediate input and output from both subnetworks in the training process. 
We refer to Fukami et al.~\cite{FFT2023_survey} and a sample code (\url{http://www.seas.ucla.edu/fluidflow/codes.html}) for further details on the present machine-learning model.
In this study, model ${\cal F}$ is trained such that weights ${\bm w}$ are optimized through
\begin{align}
    {\bm w}^* = {\rm argmin}_{\bm w} ||{\bm q}_{\rm HR}-{\cal F}({\bm q}_{\rm LR};{\bm w})||_2.
\end{align}
{While this study uses an $L_2$ norm for optimization, one can consider incorporating the knowledge from the governing equations into the cost function to better constrain the solution space~\cite{raissi2019physics,FFT2023_survey}.}

% \vspace{-2mm}
\section{Results}
\label{sec:res}
% \vspace{-10mm}

\subsection{Example 1: two-dimensional decaying homogeneous isotropic turbulence}

\begin{figure}
    \centering
    \includegraphics[width=0.85\textwidth]{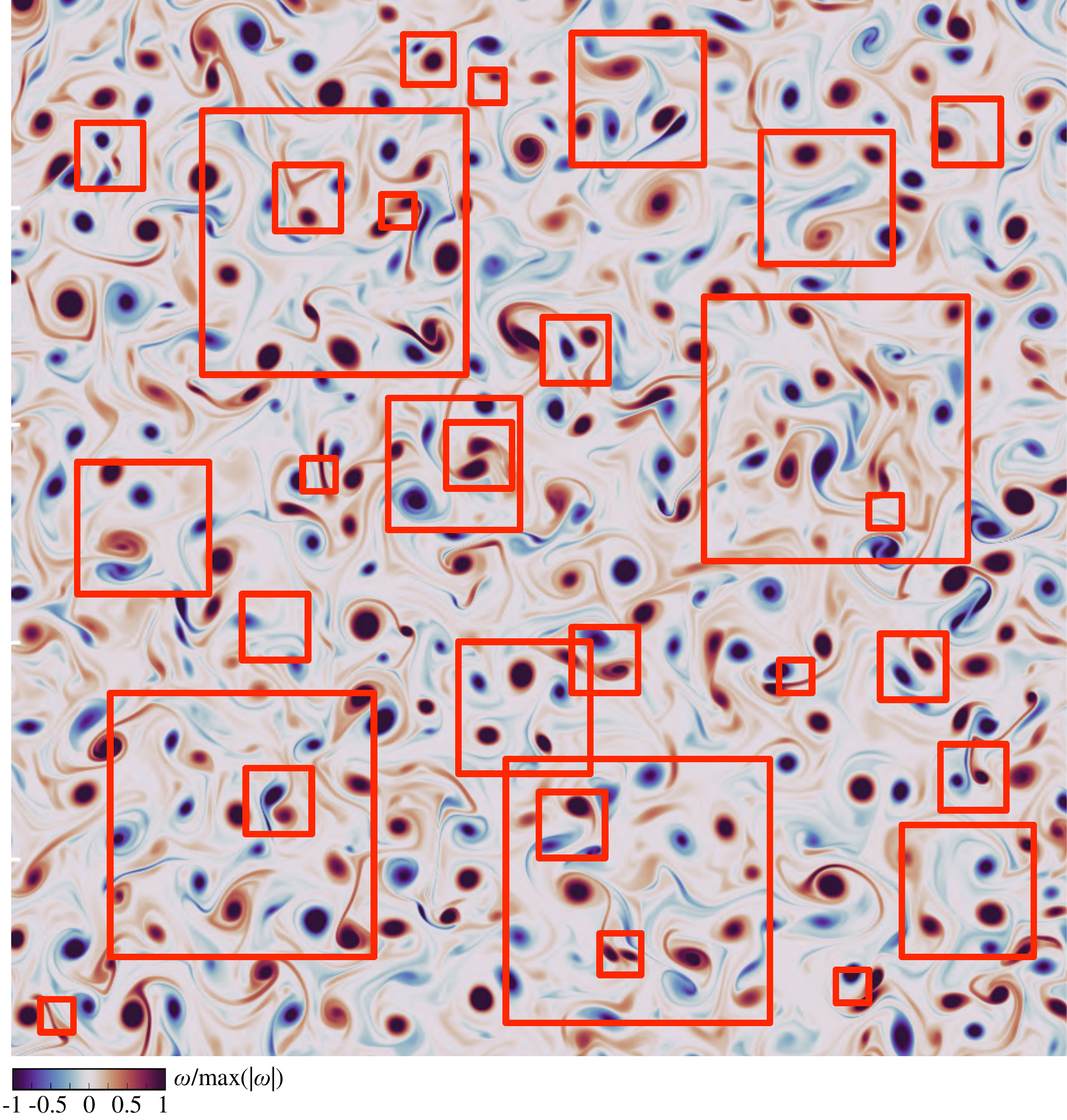}
    % \vspace{-5mm}
    \caption{
    Two-dimensional isotropic vorticity field.  
    Red boxes are example flow tiles used for training.
    }
    % \vspace{-3mm}
    \label{fig1}
\end{figure}

{
Two-dimensional decaying isotropic turbulence is {first} considered in the present single-snapshot super-resolution analysis.
The present machine-learning model is trained with subdomains collected from a single snapshot and then assessed with test snapshots obtained by independent simulations. 
The flow field data are generated with direct numerical simulation~\cite{TNB2016} that numerically solves the two-dimensional vorticity transport equation
\begin{equation}
\dfrac{\partial\omega}{\partial t}+{\bm u}\cdot\nabla\omega=\dfrac{1}{Re_0}\nabla^2 \omega,
\label{eq_1}
\end{equation} 
where ${\bm u}=(u,v)$ represents the velocity field and $Re_0=u^*l_0^*/\nu$ is the initial Reynolds number.
Here, $u^*$ is the characteristic velocity defined as the square root of the spatially averaged initial kinetic energy, {$l_0^*=[2{\overline{u^2}}(t_0)/{\overline{\omega^2}}(t_0)]^{1/2}$} is the initial integral length, and $\nu$ is the kinematic viscosity.
The overline denotes the spatial average.
The computational domain is a biperiodic square with length $L=1$.
We use the vorticity field $\omega$ as a data attribute in the present super-resolution analysis.
}

{
The baseline super-resolution analysis is performed with the model trained with a single snapshot shown in figure~\ref{fig1} with $Re_0=1580$.
Various vortical structures, including counter-rotating and co-rotating vortices and shear layers, of different length scales are contained in this single snapshot.
The number of computational grid points $N^2$ is set to~$1024^2$, satisfying $k_{\rm max}\eta \geq 1$, where $k_{\rm max}$ is the maximum wavenumber and $\eta$ is the Kolmogorov length scale, to ensure that the DNS resolves all flow scales.
The simulation for training data preparation is initialized with a distribution composed of randomly-placed Taylor vortices~\cite{taylor1918dissipation} with random strengths and sizes.
The snapshot is collected after the flow reaches the decaying regime.
}

The present training data is comprised of square-sized subdomain samples randomly collected from the single snapshot with four different sizes of $L_{\rm sub}=\{0.03125, 0.0625, 0.125, 0.25\}$, as illustrated in figure~\ref{fig1}.
The subdomain data are then resized to be $N^2_{\rm ML} = 128^2$ for the present data-driven analysis.
The dependence of super-resolution reconstruction on the choice of a single snapshot is examined later.

Test snapshots in this study are prepared from three different simulations. 
The initial Reynolds numbers and the number of grid points are, respectively, $Re_0=\{80.4, 177, 442\}$ and $N=\{128, 256, 512\}$, satisfying $k_{\rm max}\eta \geq 1$.
These settings are intended to generate test snapshots that include a similar size of vortical structures to that in the subdomains of the single snapshot.

\begin{figure}[H]
    \centering
    \includegraphics[width=\textwidth]{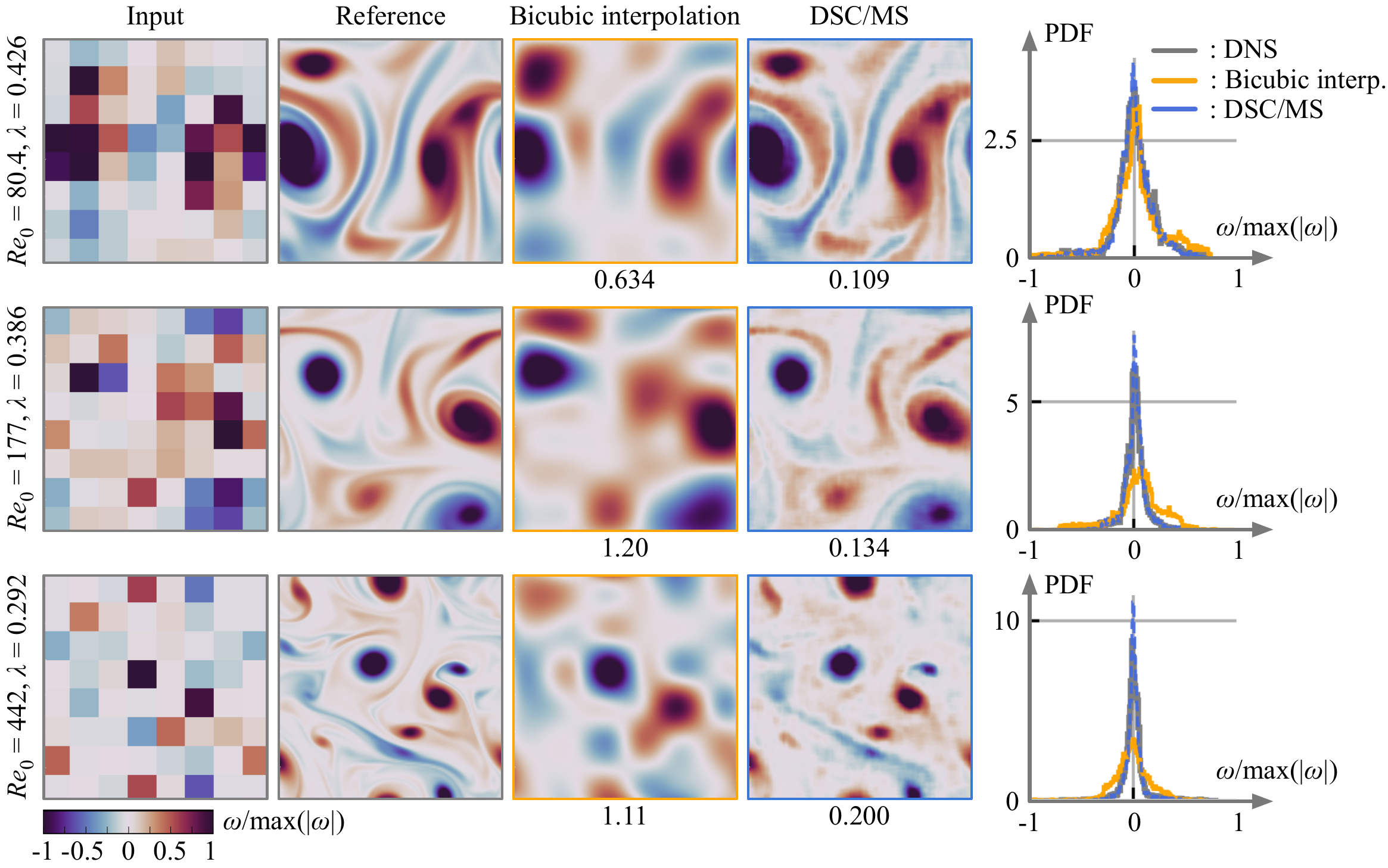}
    % \vspace{-6mm}
    \caption{
    The single snapshot super-resolution of two-dimensional decaying turbulence.
    Its accuracy is assessed with test snapshots from three different simulations.
    The instantaneous Taylor length scale $\lambda(t)$ for a representative test snapshot is reported along with each $Re_0$.
    The value underneath each contour is the $L_2$ error norm.
    The probability density function (PDF) for test snapshots at each Reynolds number is also shown.
    }
    % \vspace{-3mm}
    \label{fig3}
\end{figure}

Once the test snapshots are collected from the simulations, they are resized to be $N_{\rm ML} = 128$.
The present machine-learning model ${\cal F}$ reconstructs the high-resolution vorticity flow field of size $128^2$ from the corresponding low-resolution data of size~$8^2$ generated by average pooling~\cite{FFT2019a}. 
The input and output data are normalized by the instantaneous maximum value of absolute vorticity, ${\rm max}(|\omega|)$ to account for the magnitude difference of vorticity fields across the Reynolds number.

{

We apply the super-resolution model trained with a single snapshot to decaying turbulence at three different test $Re$.
The reconstruction by the DSC/MS model is compared to bicubic interpolation, as shown in figure~\ref{fig3}.
Let us first use 2000 local tiles in total for training the baseline model.
The value listed underneath each figure reports the $L_2$ error norm $\varepsilon = ||{\omega}_{\rm HR}-{\cal F}({\omega}_{\rm LR})||_2/||{\omega}_{\rm HR}||_2$.
As the bicubic interpolation simply smooths the given low-resolution data, the reconstructed fields do not provide any fine-scale information, resulting in a high $L_2$ error.

To improve the reconstruction of fine-scale structures, let us consider the DSC/MS model-based super resolution.
The reconstructed fields by the DSC/MS model show improved agreement with the reference data.
In addition to large-scale structures, rotational and shear-layer structures are also well represented compared to bicubic interpolation, reporting only 10-20\% $L_2$ error across the range of Reynolds numbers.
Note that this level of error suggests accurate reconstruction that captures turbulent coherent structures since the spatial $L_2$ norm is a strict comparative measure~\cite{anatharaman2023image}.

The reconstruction performance is also examined with the probability density function (PDF) of the vorticity field, as presented in figure~\ref{fig3}.
For the case of $Re_0 = 80.4$, the curves obtained from both the bicubic interpolation and the DSC/MS model are in agreement with the reference data.
However, the curve for the bicubic interpolation (colored in orange) deviates for the tail of the distribution, implying the failure in reconstructing strong rotation structures with low probability.

As the test $Re$ increases, the bicubic interpolation starts struggling to reconstruct the vorticity across its distribution.
This is because the smallest and largest scales spread wider by increasing the test Reynolds number.
In contrast, the distributions obtained by the present DSC/MS model are almost indistinguishable compared to those with the reference DNS, supporting statistically accurate reconstruction.
These results imply that even just a single turbulent flow snapshot contains a variety of vortical structures across different length scales, which can be extracted by the present machine-learning approach.

To further examine the reconstruction performance across spatial length scales, let us present in figure~\ref{fig4} the kinetic energy spectrum $E(k)$, where $k$ is the wavenumber.
While the bicubic interpolation significantly underestimates the energy across the wavenumbers, the machine-learning model provides reasonable agreement up to $k\approx 200$ for $Re_0 = 80.4$ and 177 and $k\approx 100$ for $Re_0 = 442$.
The difference in the high-wavenumber regime is due to the low correlation between the low- and high-wavenumber components, which is often observed in supervised learning-based super-resolution of turbulent flows~\cite{FFT2023_survey}.
A remedy for improved matching over the high wavenumber could be attained by using algorithms such as generative learning~\cite{kim2021unsupervised,yousif2023transformer}.
The results here indicate that the current model can learn the energy distribution over the spatial length scales and Reynolds numbers from only a single snapshot.

\begin{figure}
    \centering
    \includegraphics[width=1\textwidth]{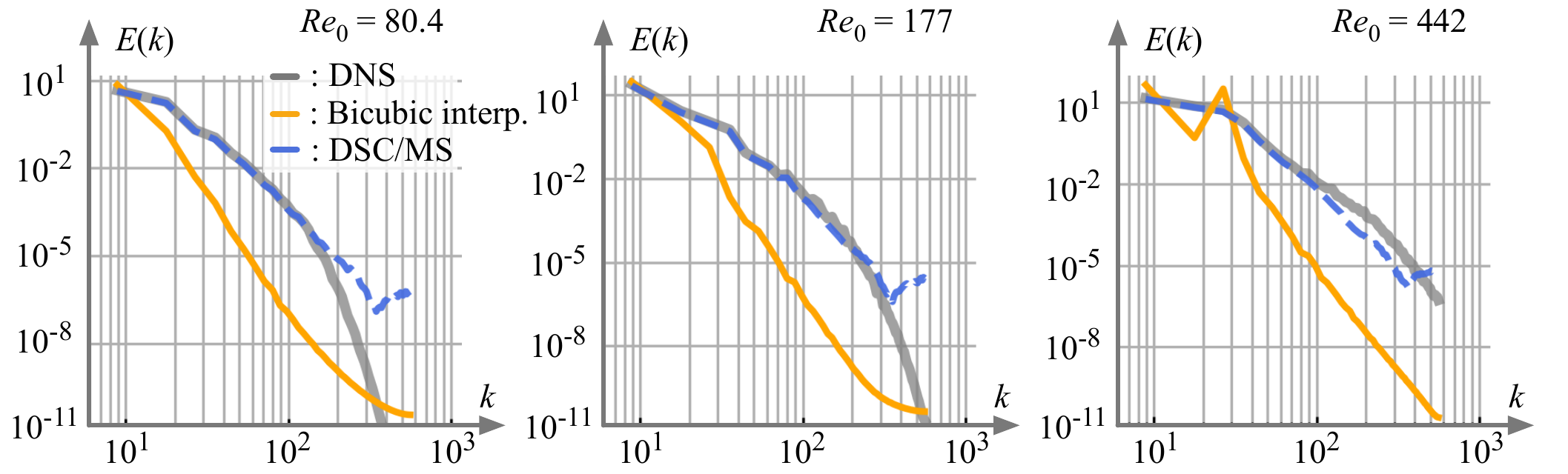}\hspace{5mm}
    \vspace{-3mm}
    \caption{
    Kinetic energy spectrum $E(k)$ of the reconstructed vorticity fields.
    }
    % \vspace{-3mm}
    \label{fig4}
\end{figure}

The successful reconstruction above is supported by the richness of vortical information contained in the training snapshot depicted in figure~\ref{fig1}.
In other words, the single snapshot to be used for training must be rich with information.
To examine this point, we further consider 150 different flow fields generated by 20 different initial conditions with $N \in [128, 2048]$ and $Re_0 \in [40, 2050]$.
We perform the single-snapshot training with these snapshots covering a variety of flow realizations regarding the size and shape of vortices, as shown in figure~\ref{fig5}.

To quantify the effect of the single-snapshot choice in training on the reconstruction performance for test data, we use the ratio of the Taylor length scale between training and test snapshots, $\lambda_{\rm test}/\lambda_{\rm single}$, where $\lambda$ represents the Taylor length scale and subscripts ``test" and ``single" denote test and training (single) snapshots, respectively.
The relationship between this ratio and the reconstruction error across the different numbers of local tiles $n_s$ generated from a vorticity snapshot is presented in figure~\ref{fig6}$(a)$.
For each case, a 3-fold cross-validation is performed and the averaged error is reported.
The reconstruction improves for large $\lambda_{\rm test}/\lambda_{\rm single}$.
In other words, large $\lambda_{\rm test}$ (low test $Re$ snapshots) or small $\lambda_{\rm single}$ (high training $Re$ snapshots) provides low reconstruction error.

\begin{figure}
    \centering
    \includegraphics[width=\textwidth]{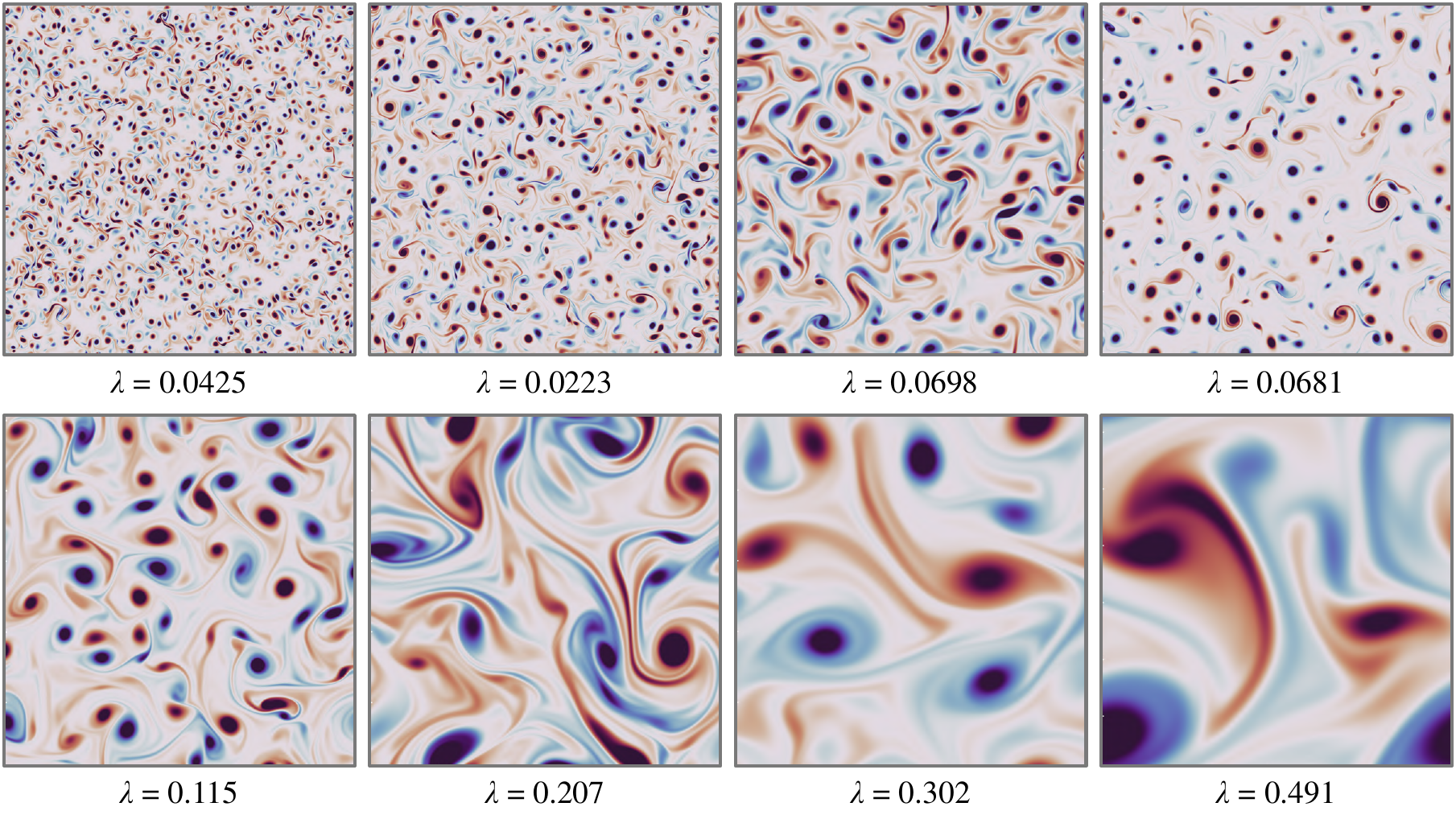}
    \vspace{-4mm}
    \caption{
    Example snapshots used for single-snapshot training.
    The value underneath each snapshot is the instantaneous Taylor length scale $\lambda(t)$.
    }
    % \vspace{-3mm}
    \label{fig5}
\end{figure}

The error decreases by increasing the number of local tiles $n_s$ across the length-scale ratio.
While this error reduction for large $n_s$ is expected, it is worth pointing out that lower $n_s$ is needed as the ratio $\lambda_{\rm test}/\lambda_{\rm single}$ increases to achieve the same level of reconstruction.
In other words, quantitative reconstruction can be achieved with a smaller number of local tiles in the single-snapshot training with a small $\lambda_{\rm single}$ that generally corresponds to a high-$Re$ field including many vortical structures.
These observations imply that in addition to the number of local samples or snapshots, the amount of information contained in the training data should also be considered when analyzing turbulent flows.

Let us focus on the baseline case of $\lambda_{\rm single} = 0.0425$, depicted in figure~\ref{fig1}, to further discuss the effect of the number of local tiles $n_s$ across the test Reynolds number, as shown in figure~\ref{fig6}$(b)$.
The averaged error over cross-validation is reported while the maximum and minimum errors at each $n_s$ are shown with shading. 
Across $n_s$, the reconstruction error at a higher $Re$ is larger compared to lower $Re$ flows, likely because of larger differences in the vortical length scales appearing in the flow.
As $n_s$ increases, the reconstruction performance is improved across the Reynolds number.
Notably, the present model achieves qualitative reconstruction for large-scale structures even with merely 250 training samples, as presented in figure~\ref{fig6}$(c)$.
Even in such a {modest} number of local tiles, there exist physical insights (relations) that can be extracted by the present super-resolution model.

Once $n_s$ exceeds 2,000, the error curves across the Reynolds number plateau, implying that extracted data of vortical flows becomes redundant from the perspective of learning. 
While the above model is trained with randomly sampled local tiles from a single snapshot, the present nonlinear machine-learning model can achieve quantitative reconstruction even with a much smaller number of local subdomains by sampling them in a smart manner based on some knowledge of the vortical flows.

The idea here is to avoid sampling local tiles that are not informative. 
To preferentially sample informative local subdomains that include insightful rotational motions and shear layers, we consider the moments of rotation and strain tensors, $W$ and $D$.
The two-dimensional probability density functions based on the mean (first moment), standard deviation (second moment, $\sigma$), skewness (third moment, $S$), and flatness (fourth moment, $F$) of $W$ and $D$ with $L_{\rm sub} = 0.0625$ are presented in figure~\ref{fig7}$(a)$.

\begin{figure}[H]
    \centering
    \includegraphics[width=0.95\textwidth]{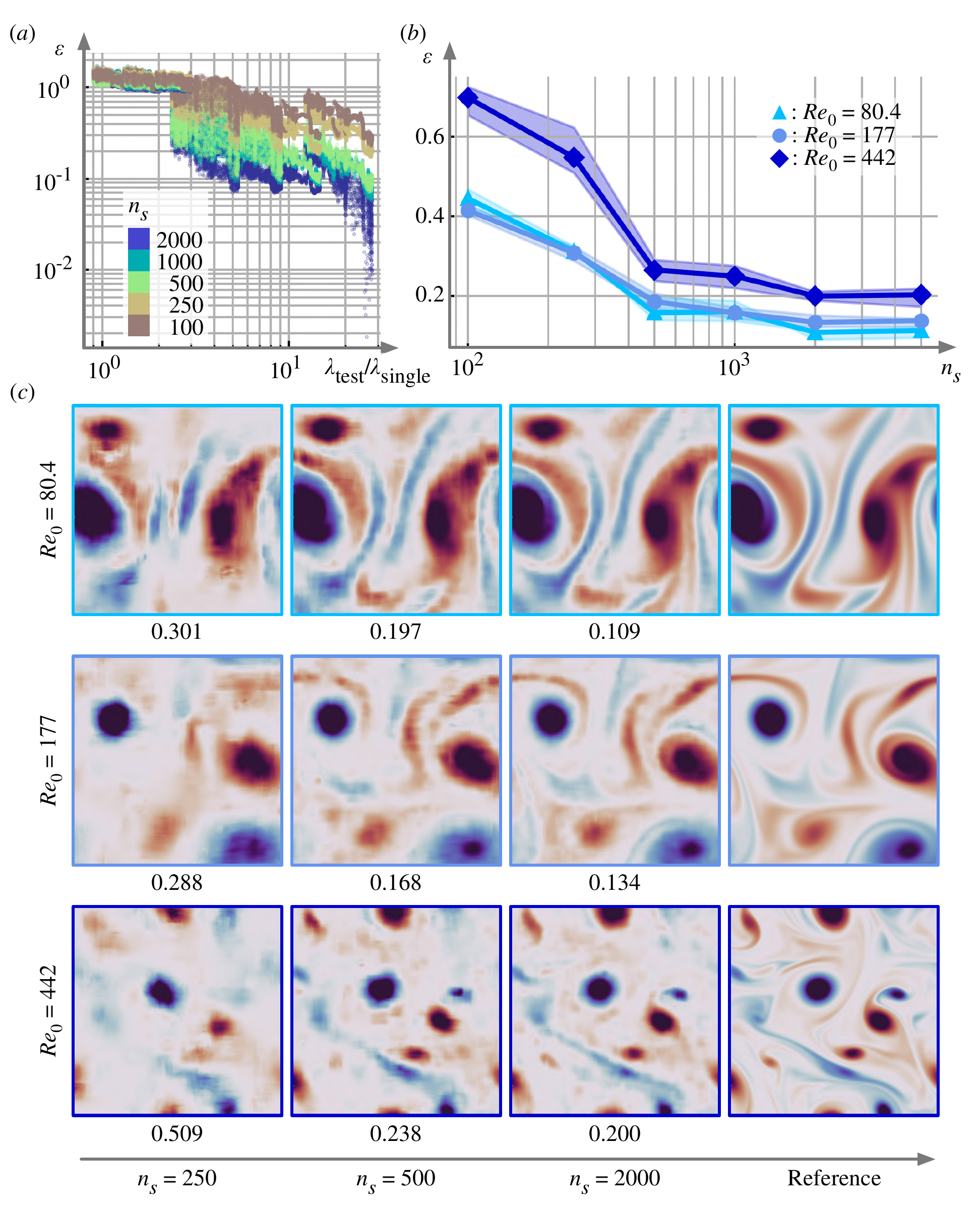}
    \vspace{-3mm}
    \caption{
    $(a)$ The relationship between the reconstruction error $\varepsilon$ and the ratio of the Taylor length scale between training and test snapshots $\lambda_{\rm test}/\lambda_{\rm single}$ across the number of training samples $n_s$.
    $(b)$ Dependence of the reconstruction performance on the number of training samples~$n_s$
    and 
    $(c)$ reconstructed vorticity fields for each test Reynolds number 
    in using a single snapshot with $\lambda_{\rm single} = 0.0425$ shown in figure~\ref{fig1}.
    The value underneath each contour in figure $(c)$ is the $L_2$ error norm.    
    }
    % \vspace{-3mm}
    \label{fig6}
\end{figure}

The 97\% confidence interval is also depicted on each PDF map.
Compared to the first and second moment-based PDFs, the third and fourth moment-based PDFs provide a sharper distribution of snapshots, as observed from the difference in the size of 97\% confidence interval area.
Furthermore, we observe that local tiles containing various structures such as flow fields (i) and (ii) appear in the region with high probability while less informative tiles such as flow fields (iii) and (iv) are seen in the area with lower probability when using the skewness.

Based on the findings above, data sampling informed by the moment probability for single-snapshot training with $n_s=250$ is performed, as shown in figure~\ref{fig7}$(b)$.
For comparison, the first and second moment-based sampling are also considered.
While the lower-order moment-based training presents similar reconstruction performance to the case in which the location of subdomains is randomly determined, the third and fourth moment-based sampling models provide enhanced reconstruction with the same number of local tiles, revealing vortices and shear-layer structures with finer details.
Note that the error level of the higher-order moment-based sampling with $n_s=250$ becomes the same as that of random sampling with $n_s=2000$, achieving significant reduction in the required number of training subsamples for accurate reconstruction.
These observations suggest that machine-learning-based analyses traditionally recognized as expensive, data-hungry approaches can take advantage of the scale-invariant property in analyzing turbulent vortical flows from much smaller data sets.

\begin{figure}
    \centering
    % \hspace{-7mm}
    \includegraphics[width=0.95\textwidth]{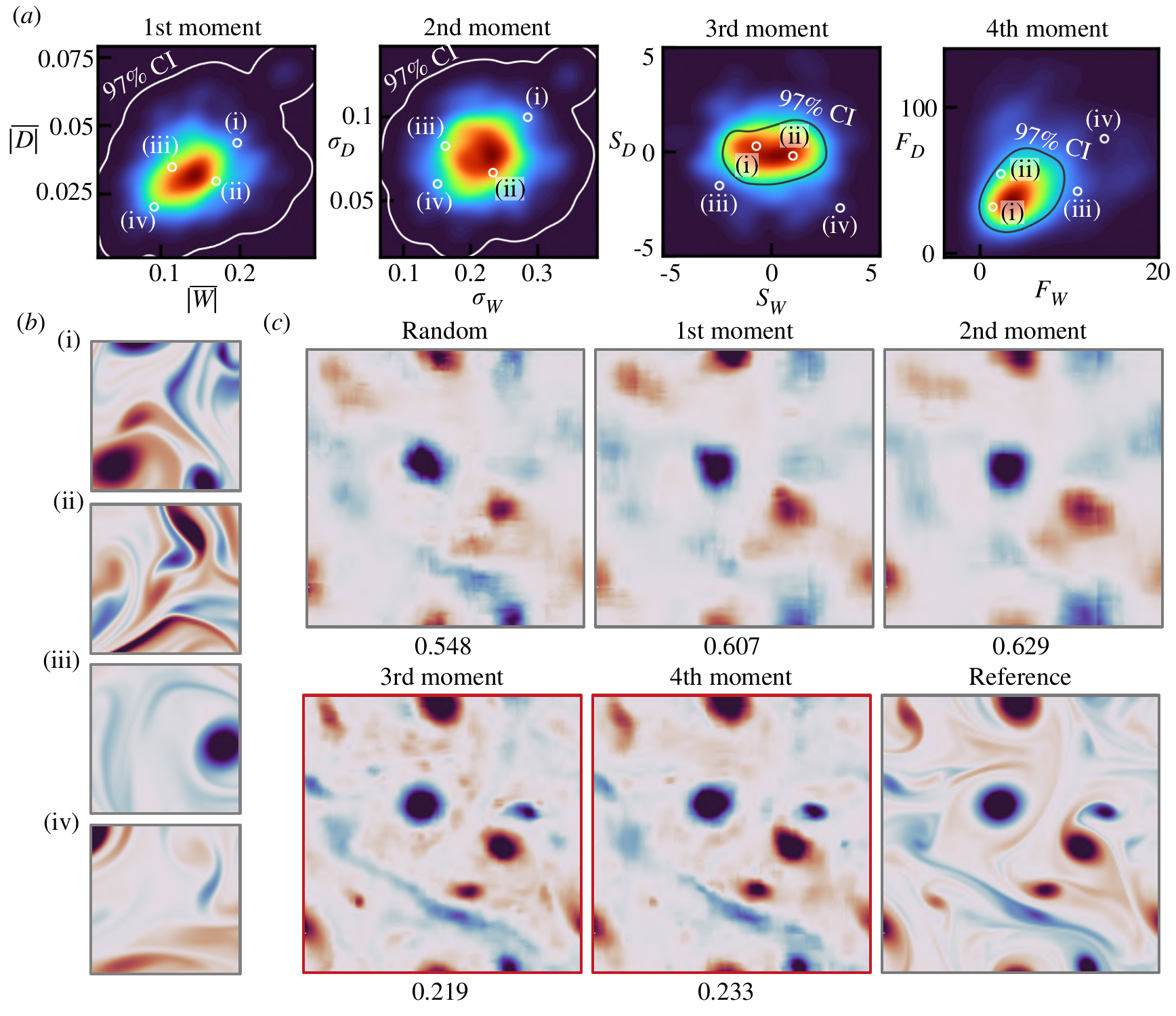}
    % \vspace{-2mm}
    \caption{
    PDF-based sampling for single-snapshot training.
    $(a)$ Two-dimensional PDF of first to fourth moments of rotation and strain tensors with $L_{\rm sub} = 0.0625$.
    For each PDF map, 97\% confidence interval is shown.
    $(b)$ Example local tiles corresponding to (i-iv) on each PDF map.
    $(c)$ Reconstruction with different data sampling with $n_s=250$.
    The value underneath each contour reports the $L_2$ error norm.
    }
    \label{fig7}
\end{figure}

{
\subsection{Example 2: turbulent channel flow}

Next, we perform the single snapshot-based super-resolution analysis for turbulent channel flow as a test case that holds spatial inhomogeneity.
For the present analysis, we consider the DNS data set made available from the Johns Hopkins Turbulence Database~\cite{perlman2007data}.
Similar to the case of two-dimensional homogeneous turbulence, the present model is trained with a collection of subdomains sampled from a single high-$Re$ snapshot and then evaluated with test snapshots obtained by an independent simulation.
The current setting enables assessing whether the present model learns flow features of turbulent channel flow across the Reynolds number from a single snapshot.

The single snapshot used for training is produced at a very high friction Reynolds number $Re_\tau = u_\tau \delta/\nu$ of 5200, holding a range of length scales~\cite{lee2015direct}.
The variables are normalized by the half-channel height $\delta$ and the friction velocity at the wall $y=0$, $u_\tau=(\nu {dU}/{dy}|_{y=0})^{1/2}$, where $U$ is the mean velocity.
The size of the computational domain and the number of grid points are $(L_{x}, L_{y}, L_{z}) = (8\pi\delta, 2\delta, 3\pi\delta)$ and $(N_{x}, N_{y}, N_{z}) = (10240, 1536, 7680)$, respectively.
Details of the numerical simulation setup are provided in Lee and Moser~\cite{lee2015direct}.

\begin{figure}
    \centering
    \includegraphics[width=0.9\textwidth]{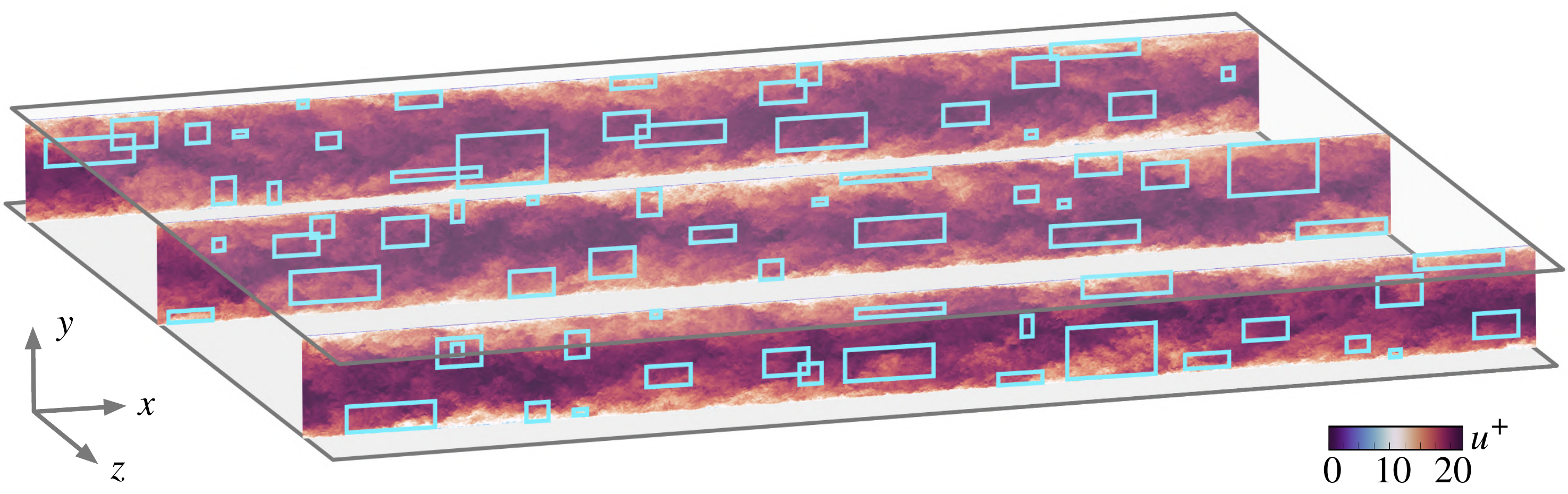}
    % \vspace{-5mm}
    \caption{
    Streamwise velocity field of turbulent channel flow at $Re_\tau = 5200$.  
    Blue boxes are example flow tiles used for training.
    }
    % \vspace{-3mm}
    \label{fig8}
\end{figure}

We consider an $x$-$y$ sectional streamwise velocity field $u$ as the variable of interest.
The subdomains used for training are randomly sampled from the $x$-$y$ sectional fields at random spanwise locations, as illustrated in figure~\ref{fig8}.
Four different sizes of subdomains are considered in the streamwise direction $L_{x,{\rm sub}}^+ = \{814, 1628, 3257, 6514\}$, where the variables with superscript $+$ denote quantities in the wall unit. 
The subdomain size in the wall-normal direction varies as the data are collected from a non-uniform grid.  
The minimum and maximum heights of the subdomains are $({\rm min}(L_{y,{\rm sub}}^+), {\rm max}(L_{y,{\rm sub}}^+)) = (66.3, 2646)$, respectively.
These collected data are resized to be $N^2_{\rm ML} = 128^2$ for the present data-driven analysis.

Test snapshots are prepared from a different DNS at $Re_\tau = 1000$~\cite{graham2016web}, also available from the Johns Hopkins Turbulence Database. 
The size of the computational domain and the number of grid points for $Re_\tau = 1000$ are $(L_{x}, L_{y}, L_{z}) = (8\pi\delta, 2\delta, 3\pi\delta)$ and $(N_{x}, N_{y}, N_{z}) = (2048, 512, 1536)$, respectively.
Details on the numerical simulation setup for this test data are available in Graham et al~\cite{graham2016web}.
The present test data are randomly subsampled from the $x$-$y$ sectional streamwise velocity field at random spanwise locations.
The super-resolution model is trained to reconstruct the high-resolution velocity field of size $128^2$ from the corresponding low-resolution data of size~$8^2$ generated by average pooling~\cite{FFT2021}.
The input and output data of streamwise velocity fields are normalized by the friction velocity $u_\tau$ to learn a universal relation between the low- and high-resolution data of turbulent channel flow across the Reynolds number~\cite{kim2021unsupervised}.

\begin{figure}[t]
    \centering
    \includegraphics[width=0.95\textwidth]{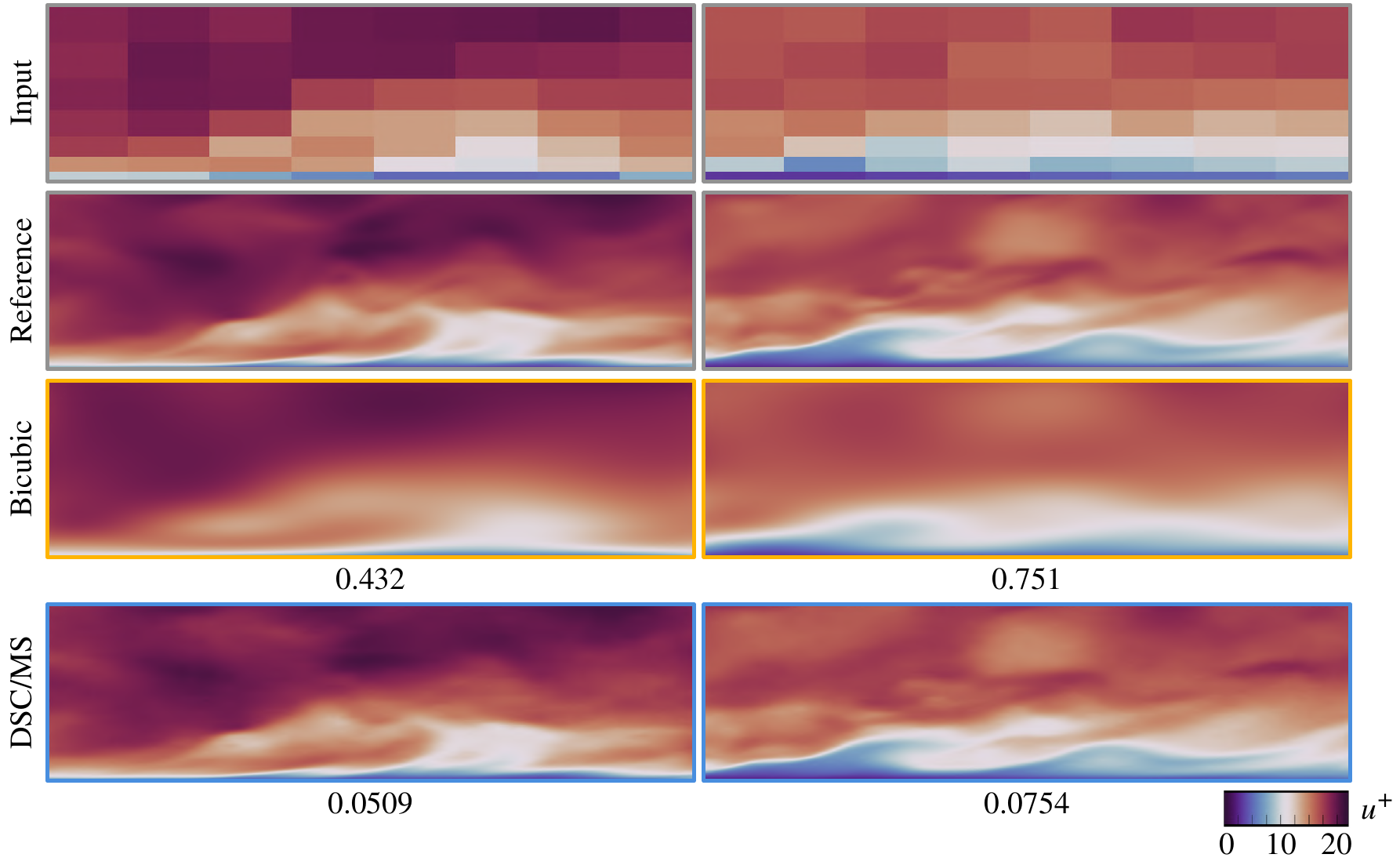}
    \vspace{-1mm}
    \caption{
    The single snapshot super-resolution analysis of turbulent channel flow.
    Its accuracy is assessed with test snapshots collected from a different simulation at $Re_\tau = 1000$.
    The value underneath each contour plot is the $L_2$ error norm normalized by streamwise velocity fluctuation.
    }
    % \vspace{-3mm}
    \label{fig9}
\end{figure}

Let us apply the present machine-learning model trained with a single snapshot at $Re_\tau = 5200$ to test datasets.
The baseline super-resolution model of the case of turbulent channel flow is trained with 2000 local tiles.
The reconstructed turbulent flow fields for two representative flow tiles of test data with $(L_{x,{\rm test}}^+, L_{y,{\rm test}}^+) = (1570, 300)$ are presented in figure~\ref{fig9}.
Here, the first grid point in the wall-normal direction of the test snapshots $y_{0, {\rm test}}$ is set to zero to examine the reconstruction performance near the wall.
To assess whether the fluctuation component of the velocity field is captured, the $L_2$ error norm in the case of turbulent channel flow reported hereafter is normalized by the velocity fluctuation such that $\varepsilon^\prime = ||{u}_{\rm HR}-{\cal F}({u}_{\rm LR})||_2/||{u}^\prime_{\rm HR}||_2$.

The reconstructed turbulent flow fields by the present machine-learning model are in agreement with the reference DNS data with as little as $5$ to $8\%$ error.
For comparison, we also show reconstruction from bicubic interpolation, which can only smooth the given low-resolution velocity fields.  
In contrast, the DSC/MS model accurately reproduces the fine-scale structures in the flow fields.

Accurate reconstruction by the present machine-learning model is also evident from statistics of the streamwise velocity field.
The DSC/MS model is superior to the bicubic method especially in reconstructing the low-speed component, as seen in the probability density function of $u^+$ shown in figure~\ref{fig10}$(a)$.
The difference in the reconstruction performance between the DSC/MS model and the bicubic interpolation is further reflected in the high-order moments depicted in figures~\ref{fig10}$(b)$, $(c)$, and $(d)$.
Note that the skewness $S(u^+)$ of the bicubic method almost matches the reference value as the low-resolution input does not hold any negative values thereby producing a distribution skewed toward positive values.
The flatness $F(u^+)$ particularly captures the difference in the produced distributions, supporting successful reconstruction by the present machine-learning model.

Turbulence statistics of the reconstructed velocity fields are also evaluated.
The root mean square of streamwise velocity fluctuation $u_{\rm rms}$ and the mean velocity profile across the wall-normal direction are presented in figures~\ref{fig11}$(a)$ and $(b)$, respectively.
The statistics obtained by the DSC/MS model accurately match the reference DNS across the $y$ direction while the bicubic interpolation produces overestimation of the velocity field at the viscous sublayer and part of the buffer layer of $y^+\lesssim 10$.

\begin{figure}[H]
    \centering
    \includegraphics[width=\textwidth]{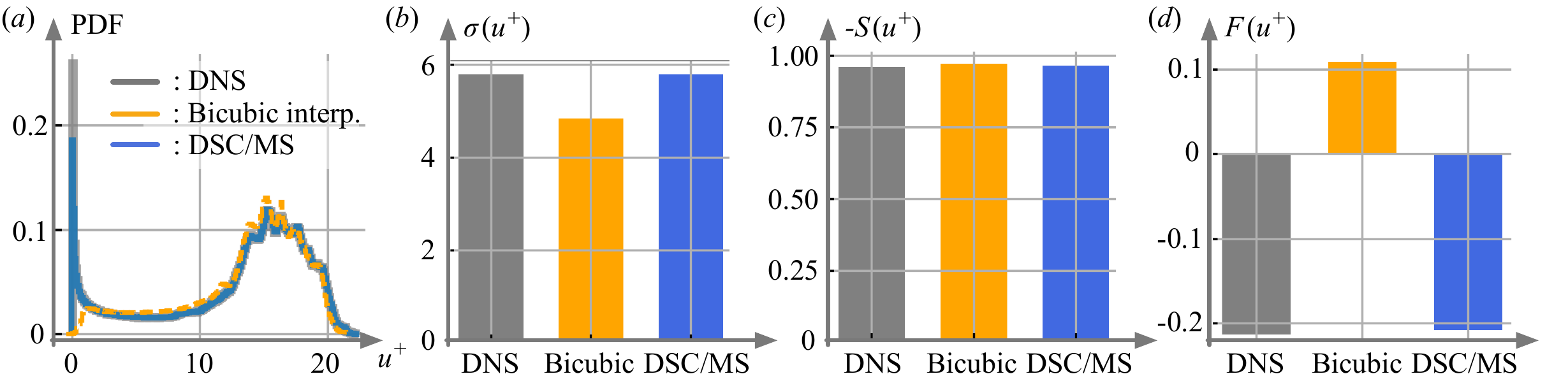}
    \vspace{-5mm}
    \caption{
    Statistics of the streamwise velocity.
    $(a)$ Probability density function (PDF), $(b)$ second, $(c)$ third, and $(d)$ fourth moments of the streamwise velocity $u^+$.
    The negative value of the third moment $S(u^+)$ is presented in figure~$(c)$. 
    }
    % \vspace{-3mm}
    \label{fig10}
\end{figure}

\begin{figure}
    \centering
    \includegraphics[width=0.8\textwidth]{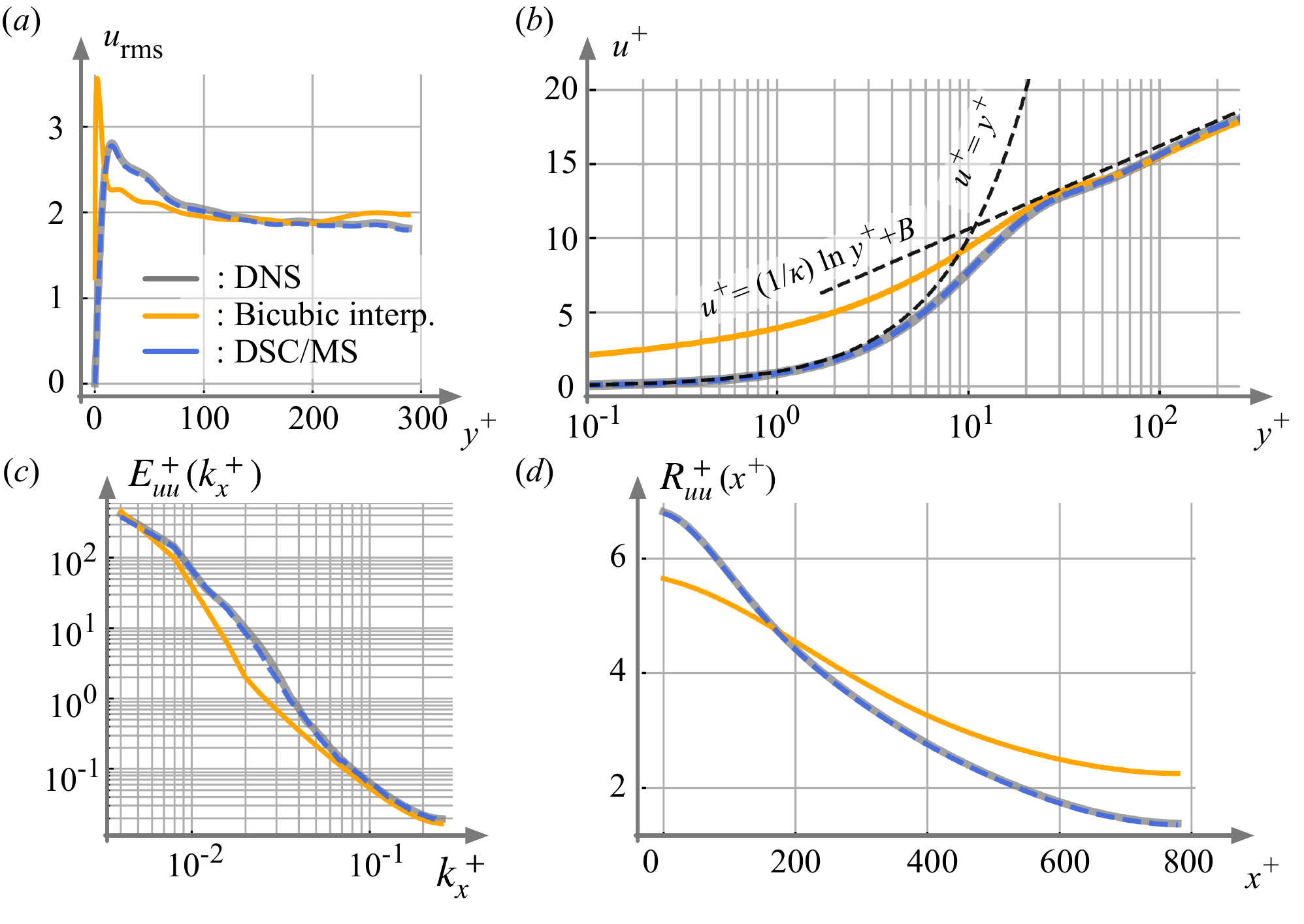}
    \vspace{-2mm}
    \caption{
    Turbulence statistics of the streamwise velocity.
    $(a)$ Root mean square of streamwise velocity fluctuation $u_{\rm rms}$.
    $(b)$ Mean velocity profile across the wall-normal direction.
    The coefficients $\kappa$ and $B$ for the logarithmic law of the wall are set to 0.41 and 5, respectively.
    $(c)$ Streamwise kinetic energy spectrum $E_{uu}^+(k_x^+)$ and $(d)$ spatial two-point correlation coefficient $R_{uu}^+(x^+)$ at $y^+=10.4$.
    }
    % \vspace{-3mm}
    \label{fig11}
\end{figure}

To further examine the reconstruction performance near the wall, we assess the streamwise kinetic energy spectrum $E_{uu}^+(k_x^+)$, where $k^+_x$ represents the streamwise wavenumber, and the spatial two-point correlation coefficient $R_{uu}^+(x^+)$ at $y^+=10.4$, depicted in figures~\ref{fig11}$(c)$ and $(d)$, respectively.
The energy distribution across the wavenumber is well represented with the DSC/MS model.
In addition, the decaying profile of the spatial two-point correlation coefficient over $x^+$ is accurately reproduced by the present machine-learning model, suggesting that the streamwise flow pattern in a flow field is super-resolved well with the DSC/MS model.
These results imply that features of turbulent channel flow across $Re_\tau$~\cite{reynolds1967stability,yamamoto2018numerical} can be successfully extracted by nonlinear machine learning from only a given single snapshot.

\begin{figure}[H]
    \centering
    \includegraphics[width=\textwidth]{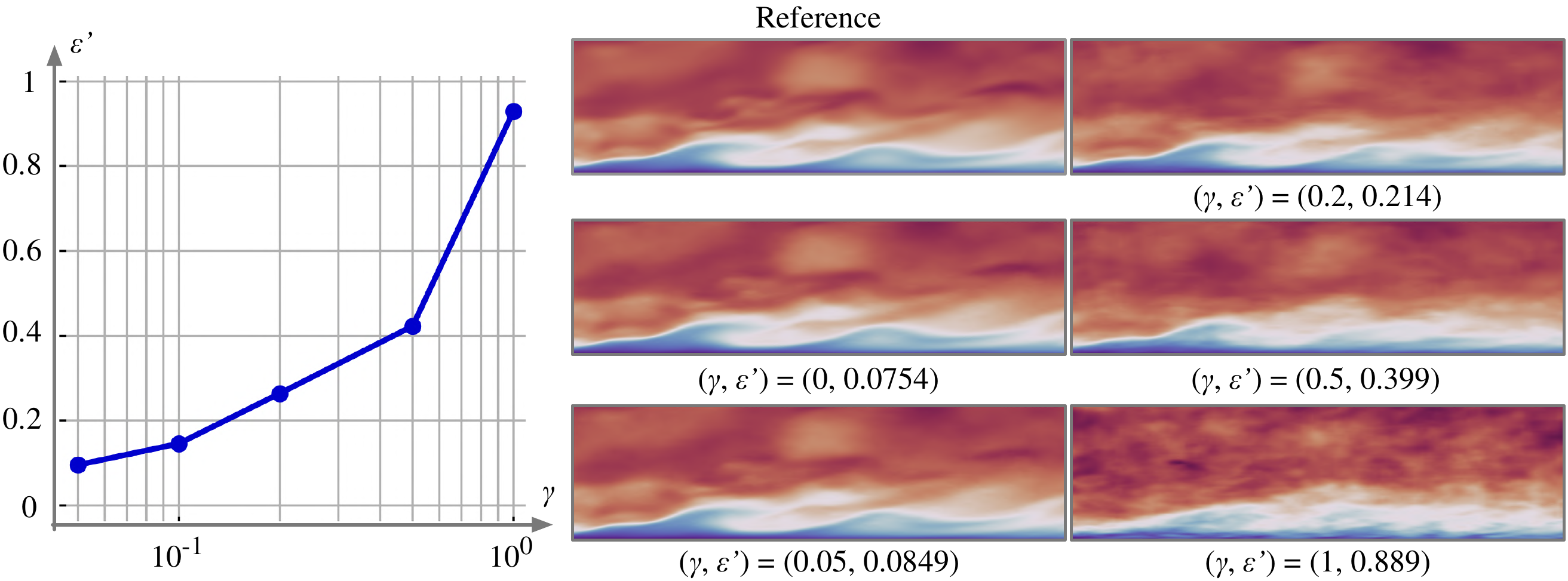}
    \vspace{-3mm}
    \caption{
    Robustness of the machine-learning model trained with 2000 subsamples against noisy low-resolution flow field input.
    The magnitude of noise $\gamma$ and the $L_2$ error norm normalized by streamwise velocity fluctuation are shown underneath each contour.}
    % \vspace{-3mm}
    \label{fig12}
\end{figure}

Here, let us assess the effect of input noise on single-snapshot super-resolution reconstruction.  
For the present assessment, the Gaussian noise $\bm n$ is given to a low-resolution input such that ${\bm q}_{\rm LR,noise} = {\bm q}_{\rm LR}+{\bm n}$, where the magnitude of noisy input $\gamma$ is given as $\gamma = ||{\bm n}||/||{\bm q}||$.
The relationship between the noise magnitude and the $L_2$ error norm normalized by streamwise velocity fluctuation $\varepsilon^\prime$ is depicted with representative reconstructed fields in figure~\ref{fig12}.
The error increases with the magnitude of noise $\gamma$.  
While the present model accurately reconstructs fine structures in a flow field up to $\gamma\approx 0.2$, large-scale structures can be reconstructed even with 50\% noise, exhibiting reasonable robustness for the given noise levels.

At last, the dependence of super-resolution reconstruction on the number of local tiles $n_s$ for the turbulent channel flow is examined, as shown in figure~\ref{fig13}.
The averaged fluctuation-based error over three-fold cross-validation is shown with the maximum and minimum errors at each $n_s$ indicated by the shading. 
The reconstruction performance improves monotonically with increasing $n_s$.
While the region far away from the wall is reasonably reconstructed with $n_s = 500$, more subdomains with the order of $O(10^3)$ are required for accurate reconstruction near the wall, likely because of the difference in flow complexities across the wall-normal direction of turbulent channel flow.
In turn, these results suggest that nonlinear machine learning can extract physical insights of turbulent flows even with spatial inhomogeneity by sufficiently collecting training subsamples from a {\it single} turbulent flow snapshot.
}

\vspace{-4mm}
\section{Concluding remarks}
\label{sec:conc}

This study discussed how we can efficiently extract physical insights from a very limited amount of turbulent flow data with machine learning.
We considered machine-learning-based super-resolution reconstruction with training data of a single turbulent flow snapshot, enabling the evaluation of whether a physical relationship between high- and low-resolution flow fields can be learned from limited available flow data.
A convolutional network-based super-resolution model, the DSC/MS model, is trained with local flow subdomains collected from only a single turbulent flow snapshot and then assessed for test data generated from different simulations.
With an example of two-dimensional decaying isotropic turbulence, we showed that training data for super-resolution analysis can be efficiently prepared from a single flow snapshot based on their statistical characteristics.
We also performed the single snapshot-based super-resolution for turbulent channel flow, showing that it is possible to learn physical relations between low- and high-resolution flow fields from a single snapshot even with spatial inhomogeneity.

\begin{figure}
    \centering
    \includegraphics[width=\textwidth]{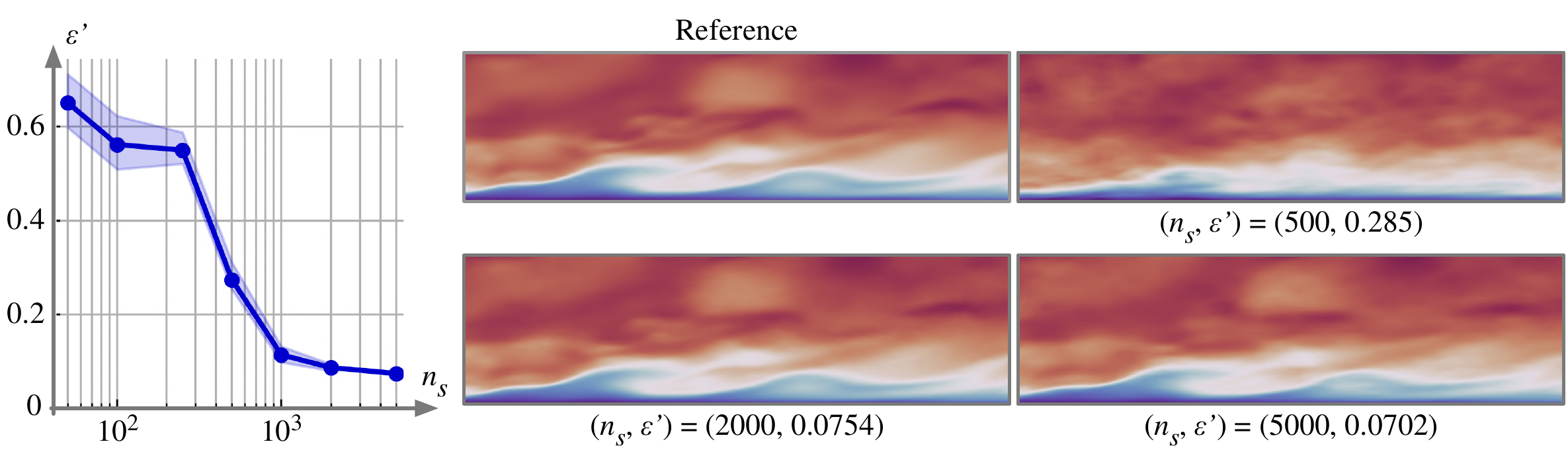}
    \vspace{-3mm}
    \caption{
    Dependence of the reconstruction performance on the number of training subsamples~$n_s$ for turbulent channel flow.
    $n_s$ and the $L_2$ error norm normalized by streamwise velocity fluctuation are shown underneath each contour.
    }
    % \vspace{-3mm}
    \label{fig13}
\end{figure}

Although machine-learning-based analysis is often described as data-intensive, our findings indicate that it is possible to extract physical insights without over-relying on massive training data for studying turbulent flows.
Capturing universal flow features across the Reynolds number such as scale-invariant characteristics is the key to successful turbulent flow reconstruction with data-driven techniques.
The use of an appropriate model architecture with physics embedding depending on flows of interest is important.
The current results also imply that redundancy of turbulent flows in not only space but also time can also be considered in sampling training data. 
By incorporating prior knowledge for developing a machine-learning model and collecting training data, we should be able to use smaller data sets to learn physics in a much smarter manner.
We should stop being wasteful of turbulent flow data.
}

% \vspace{-1.5mm}
\section*{Acknowledgements}
We thank the support from the US Air Force Office of Scientific Research (FA9550-21-1-0178) and the US Department of Defense Vannevar Bush Faculty Fellowship (N00014-22-1-2798).
The computations for machine-learning analysis were performed on Delta GPU at the National Center for Supercomputing Applications (NCSA) through the ACCESS program (Allocation PHY230125).

% \vspace{-1.5mm}
\section*{Declaration of interests}

{The authors report no conflict of interest.}

%%%%%%%%%%%%%%%%%%%%%%%%%%%%%%%%
%%%%%%%%%%%%%%%%%%%%%%%%%%%%%%%%%
%%%%%%%%%%%%%%%%%%%%%%%%%%%%%%%%%

\bibliographystyle{unsrt}  
\bibliography{refs}

\end{document}